\documentstyle[12pt,epsf]{article}
\newcommand{\la}[1]{\label{#1}}
\newcommand{\lsim}{{\buildrel < \over {_\sim}}\,}
\newcommand{\gsim}{{\buildrel > \over {_\sim}}\,}

\begin{document}
\begin{centering}
{\large\bf  MINIJETS IN ULTRARELATIVISTIC HEAVY ION COLLISIONS 
AT RHIC AND LHC}
\vspace{0.5cm}

{\large K.J.Eskola}\footnote{On leave of absence from: 
{\it Laboratory of High Energy Physics, Department of Physics,
P.O.Box 9, 00014 University of Helsinki, Finland}}

{\it CERN/TH, CH-1211 Geneve 23, Switzerland}

\end{centering}
\begin{abstract}

\noindent Recent results on minijet production in nuclear collisions 
at the RHIC and LHC energies are reviewed. Initial conditions of the 
QGP at $\tau=0.1$ fm$/c$, especially parton chemistry, thermalization 
and net baryon number-to-entropy ratio are discussed. Also, contribution 
of minijets from a hard BFKL-pomeron ladder will be estimated.
\end{abstract}
\section{Introduction} 

Particle and transverse energy production in the central rapidity 
region of heavy ion collisions can be treated as a combination of 
hard/semihard parton production  and soft particle production. 
With increasing energies, the semihard  QCD-processes are expected 
to become increasingly important. This is due to two reasons: 
firstly, already in $p\bar p(p)$ collisions the  rapid rise 
of the total and inelastic cross sections  can be explained by 
copious production of semihard partons, {\it minijets}, with transverse 
momenta $p_{\rm T}\ge p_0\sim 1...2$ GeV \cite{EIKONAL}. This is also 
expected to happen in $AA$ collisions at very high energies.
Secondly, the semihard particle production scales as $A^{4/3}$, 
so that for large  nuclei the importance of semihard partons is 
increased further \cite{BM87,KLL,EKR}. 
The soft, non-perturbative, particle production  in ultrarelativistic  
heavy ion collisions can be modelled {\it e.g.} through strings 
\cite{FRITIOF, VENUS,HIJING} or through a decaying strong background 
colour field \cite{KM85}. 

The time scale for producing partons  and transverse energy into 
the  central rapidity region by semihard collisions is short, typically  
$\tau_{\rm h}\sim 1/p_0\sim 0.1$ fm$/c$, where $p_0\sim 2$ GeV 
is the smallest  transverse momentum included in the computation. 
The soft processes are completed  at later stages of the collision, at 
$\tau_{\rm s}\sim 1/\Lambda_{\rm QCD}\sim 1$ fm$/c$. 
If the density  of partons produced in the hard and semihard stages 
of the heavy ion collision becomes high enough - as will be the case 
- a saturation in the initial parton production can occur 
\cite{BM87,GLRpr,EMW,EK96}, and softer particle production will be 
screened. The fortunate consequence of this is that a larger part of 
parton  production in the central rapidities can be {\it computed} 
from perturbative QCD (pQCD) at higher energies and the relative 
contribution from soft collisions with $p_{\rm T}\lsim 2$ GeV becomes 
smaller. Typically, the expectation is that 
at the SPS (Pb+Pb at $\sqrt s=17$ $A$GeV),  the soft component dominates, 
and at the LHC (Pb+Pb at $\sqrt s=5.5$ $A$TeV) the semihard component 
is the dominant one. At the RHIC (Au+Au at $\sqrt s=200$ $A$GeV) one 
will be in the intermediate  region, and both components should be 
taken into account. 

A lot of effort has also been devoted for building event generators 
\cite{HIJING,PCM} addressing the dominance of semihard processes in 
nuclear collisions at high energies. These have generated also new 
insight and very useful discussion during the recent years. 
Also recently, a promising novel approach to minijet production 
has been developed \cite{McL}.

I have divided this talk basically into two halves. In the first one, I will 
recapitulate the basic features of semihard parton production and review 
our latest results \cite{EKR,EMW,EK96}. The main goal of these studies is to 
find out the initial conditions for early QGP-formation at 
$\tau\sim 0.1$ fm$/c$, 
including the transverse energy deposited into the mid-rapidity region, 
chemical composition of the parton plasma, and, to
study the possibility of a very rapid thermalization and estimate the 
initial net baryon-to-entropy ratio. It is vitally important to study 
the early formation of strongly interacting partonic 
matter, since the later evolution of the QGP, 
the final state global observables, and the suggested signals of the 
plasma will strongly depend on the initial conditions.
The second half I will devote for discussion of an additional mechanism
for parton and transverse energy production: minijets from a BFKL-ladder
\cite{ELR96}. Especially, I will estimate the maximum 
amount of transverse energy one should  expect from the BFKL-minijets
in heavy ion  collisions.

\section{Initial conditions for QGP at $\tau\sim 0.1$ fm$/c$}

Hadronic jets originating from high $p_{\rm T}$ quarks and gluons 
are clearly observed experimentally but when the partons 
have $p_{\rm T}\lsim 5$ GeV  the jets become very difficult to 
distinguish  \cite{UA188} from the underlying event.  
In heavy ion collisions, where we expect hundreds (RHIC) or thousands 
(LHC) of minijets with  $p_{\rm T}\sim 2$ GeV be produced, detection 
of individual minijets will be impossible. However, the semihard partons 
are expected to contribute dramatically to the early formation of QGP. 
The idea of multiple production of semihard gluons and quarks in $pp$ 
and $AA$ collisions is based on a picture of independent binary 
parton-parton collisions. The key quantity is the integrated 
jet cross section,
\begin{equation}
\sigma_{\rm jet}(\sqrt s, p_0) = 
\frac{1}{2}\int_{p_0^2} dp_T^2dy_1dy_2\sum_{{ijkl=}\atop{q,\bar q,g}}
        \int dy_2x_1f_{i/N}(x_1,Q) x_2f_{j/N}(x_2,Q)
        {d\hat\sigma\over d\hat t}^{ij\rightarrow kl}
        {\hskip-7mm}(\hat s,\hat t,\hat u),
\label{sigmajet}
\end{equation}
where $x_{1,2}$ are the fractional momenta of the incoming partons 
$i$ and $j$, and $f_{i/N}(x,Q)$ are the parton distributions in $N$ 
($=p,A$). The factor 2 comes from the fact that, in the lowest 
order (LO) pQCD, there are two partons produced in each semihard 
subcollision. In the eikonal models for $pp$ collisions \cite{EIKONAL} 
the ratio $\sigma_{\rm jet}/\sigma_{\rm inelastic}$ can be interpreted 
as the average number of semihard events in one inelastic 
collision. The results I will be quoting in the following \cite{EKR} are 
obtained with the MRSH \cite{MRSH} and MRSD-' \cite{MRSD} 
parton distributions with a scale choice  $Q=p_{\rm T}$. 
More detailed formulation can be found in Refs. \cite{KLL,EK96},
and numerical evaluation of Eq. (\ref{sigmajet}) in Ref. \cite{EKR}.

The formula above is defined in the lowest order (LO), 
$d\hat\sigma/d\hat t\sim\alpha_{\rm s}^2$. Often a constant factor 
$K\sim 2$ is used to simulate the effects of NLO terms. 
Studies of the NLO jet cross section $d\sigma/(dp_{\rm T}dy)$ \cite{EKS}
show that (with a scale choice $Q=p_{\rm T}$ and with a jet size $R\sim 1$)
 this is a reasonable approximation \cite{EWatHPC}. Strictly speaking, 
however, a theoretical $K$-factor can only be defined for quantities 
where a well-defined, infrared-safe measurement function can be applied 
\cite{EKS}. For $E_{\rm T}$-production in nuclear collisions, an 
acceptance window in the whole central rapidity unit defines such a 
function but for this acceptance criteria and for $p_{\rm T}\sim 2$ GeV 
the exact NLO contribution has not been computed yet. 

The first estimate of the average number of produced semihard partons with 
$p_{\rm T}\ge p_0$ in an $AA$ collision at a fixed impact parameter ${\bf b}$
can be obtained as \cite{KLL}
\begin{equation}
\bar N_{AA}({\bf b},\sqrt s,p_0) = 
2T_{AA}({\bf b})\sigma_{\rm jet}(\sqrt s,p_0),
\label{N}
\end{equation}
and the average transverse energy carried by these partons as \cite{KLL}
\begin{equation}
\bar E_{\rm T}^{AA} ({\bf b},\sqrt s,p_0) = T_{AA}({\bf b})\sigma_{\rm jet}(\sqrt s,p_0)\langle E_{\rm T}\rangle,
\label{ET}
\end{equation}
where $T_{AA}({\bf b})$ is the nuclear overlap function \cite{KLL} 
which scales $T_{AA}\sim A^{4/3}$, describing thus the typical scaling 
of hard processes in nuclear collisions. The normalization is
$\int d^2{\bf b} T_{AA}({\bf b}) = A^2$ and, for large nuclei 
with Woods-Saxon nuclear densities, 
$T_{AA}({\bf 0})\approx A^2/(\pi R_A^2)$. 
The acceptance criteria imposed for the quantities 
$\sigma_{\rm jet}(\sqrt s,p_0)$
and for $\sigma_{\rm jet}(\sqrt s,p_0)\langle E_{\rm T}\rangle$
will be $|y|\le 0.5$, and the corresponding cuts will be made 
in $y_1$ and $y_2$. In Eqs. (\ref{N}) and (\ref{ET}) above,
$T_{AA}({\bf b})\sigma_{\rm jet}$ is the average number of semihard collisions
and $\langle E_{\rm T}\rangle$ is the  average transverse energy carried 
by the partons produced in each of these collisions. 
We fix $p_0=2$ GeV, {\it i.e.} we describe  the initial conditions at
$\tau\sim 1/p_0=0.1$ fm$/c$. The predictions for the central rapidity unit in 
Pb-Pb collisions at the RHIC and LHC energies are summarized in Tables 1. 
Also, contributions from gluon, quark and antiquark production are shown 
separately \cite{EKR}.\footnote{Results for $d\sigma/(dp_{\rm T}dy)$ 
at $y=0$ can be  found in Ref. \cite{EMW}.}

In the results given, we have neglected nuclear effects in parton 
distributions: $f_{i/A}=Af_{i/p}$. 
In reality, however, in the typical $x$-region
$x\sim 2p_0/\sqrt s$ there are quite strong shadowing corrections
\cite{NMC}, especially for the LHC nuclear collisions. Also, the scale 
evolution of nuclear gluon shadowing was shown to be 
potentially important in the analysis in Ref. \cite{KJE}. 
However, a re-analysis with the input from HERA at small-$x$ 
\cite{HERA93} has to be performed before 
getting a solid quantitative prediction of the shadowing effects
on minijet production.

The rapid rise of the structure function $F_2(x,Q)$ at small values of $x$ 
observed at HERA \cite{HERA93} does not affect the bulk of 
the 2 GeV minijets at RHIC energies very much but obviously has quite 
dramatic consequences at the LHC energies. As demonstrated in \cite{EKR}, 
there is a clear enhancement of minijet production due to the new parton 
distributions. However, the more rapidly the gluon distributions rise, 
the more there should be nuclear shadowing due to  the GLRMQ-fusions 
\cite{GLRpr,MQ87,LEVIN}. Again, a more quantitative prediction 
depends on the scale evolution of nuclear gluon shadowing as well.

\begin{table}
\center
(a)
\begin{tabular}{|c|c|c|c|c|c|}
\hline
$\bar N_{\rm PbPb}$         & total  & $g$   & $q$    & $\bar q$   \\
\hline
LHC:                        & 3252   & 2710  & 276    & 266        \\
                            & 5978   & 5220  & 385    & 373        \\
\hline 
RHIC:                       & 200    & 156   & 26.3   & 17.4       \\      
                            & 199    & 157   & 25.7   & 16.6       \\
\hline
\end{tabular}
\hspace{1cm}
(b)
\begin{tabular}{|c|c|c|c|c|c|}
\hline
$\bar E_{\rm T}^{\rm PbPb}$ & total   & $g$   & $q$    & $\bar q$   \\
\hline
LHC                         & 10310   & 8640  & 854    & 816        \\
                            & 17580   & 15330 & 1150   & 1100       \\
\hline 
RHIC                        & 547     & 426   & 73.6   & 47.0       \\      
                            & 539     & 422   & 71.7   & 44.8       \\
\hline
\end{tabular}
\caption[1]{{\footnotesize 
{\bf (a)} The average numbers of semihard partons at $\tau=0.1$ fm$/c$
with $|y|\le 0.5$ and $p_{\rm T}\ge 2$ GeV in central Pb-Pb collisions,
as given by Eq. (\ref{N}). Shadowing is not included and $K=2$.
The upper values are obtained with MRSH and the lower ones with MRSD-' parton 
distributions.
{\bf (b)} The average transverse energy carried by these partons, 
as predicted by Eq. (\ref{ET}).  }}
\la{table1}
\end{table}

Let us now have a closer look at the results in Table 1. There are four 
important observations. Firstly, the gluons clearly dominate both the 
initial parton and transverse energy production: the initial parton 
system is about 80 \% glue. 

Secondly, the effective transverse area of the 
produced semihard partons is $\bar N_{AA} \pi/p_0^2$. Comparing this 
with the effective nuclear transverse area, $\pi R_A^2$, we notice that 
\begin{equation}
\xi_A \equiv \frac{\bar N_{AA} \pi/p_0^2}{\pi R_A^2} 
\sim 1 \,\,\,{\rm for \, LHC}
\,\,\,\,\,\,\,\,\,\,{\rm and}\,\,\,\,\,\,\,\,\,\,
\xi_A \ll 1 \,\,\,{\rm for\, RHIC},
\end{equation}
{\it i.e.} the parton system at the LHC at 0.1 fm$/c$ is already 
dense enough so that a saturation of parton production can take place 
\cite{BM87,GLRpr,EK96}. In this way, the scale $p_0$ acquires also 
{\it dynamical} significance. At RHIC, since $\xi_A<1$,
saturation occurs at smaller values of $p_{\rm T}$ (at $\tau>0.1$ fm$/c$), 
possibly  in the region where pQCD cannot be trusted.  
This qualitative argumentation is supported by a more quantitative, 
although still phenomenological, analysis of 
Ref. \cite{EMW}, where we suggested that at sufficiently large energies 
(LHC) and large nuclei ($A\sim 200$), a  dynamical screening mass is 
generated, causing a saturation in the minijet cross sections \cite{EMW} at
a perturbative scale like $p_0\sim 2\, {\rm GeV}\gg \Lambda_{\rm QCD}$. 
The consequence is that the softer parton production is screened and
its relative importance becomes smaller.\footnote{A similar saturation 
effect is also expected in the approach by McLerran {\it et al.} \cite{McL}.}

The third interesting observation is that the gluonic subsystem in the 
central rapidity unit $\Delta y = 1$ may thermalize very fast, at least 
in the LHC nuclear collisions.  In the perturbatively produced system 
the (transverse) energy per gluon is 
$\bar E_{\rm T}^g/\bar N_{\rm PbPb}^g 
= \epsilon_g^{\rm pQCD}/n_g^{\rm pQCD}\approx 3$ GeV and the energy 
density of the system at $\tau_{\rm h}=0.1$ fm/$c$ is 
$\epsilon_g^{\rm pQCD} = \bar E_{\rm T}^g/(\pi R_A^2\tau_h\Delta y)$.
The temperature $T_{\rm eq}$ of an ideal (massless boson) gas in a complete 
thermal (= both kinetic and chemical) equilibrium with this energy density 
can be computed from
$\epsilon_g^{\rm ideal} 
=3\pi^2/90\cdot 16T_{\rm eq}^4 
= \epsilon_g^{\rm pQCD}$, and we get $T_{\rm eq} =$ 0.988 (1.14) GeV 
with the MRSH  (MRSD-') densities. Especially, we find \cite{EKR}
\begin{equation}
\frac{\epsilon_g^{\rm pQCD}}{n_g^{\rm pQCD}}
\sim \frac{\epsilon_g^{\rm ideal}}{n_g^{\rm ideal}}\,\,\,{\rm for\, LHC}
\,\,\,\,\,\,\,\,\,\,{\rm and}\,\,\,\,\,\,\,\,\,\,
\frac{\epsilon_g^{\rm pQCD}}{n_g^{\rm pQCD}}
> \frac{\epsilon_g^{\rm ideal}}{n_g^{\rm ideal}}\,\,\, {\rm for\, RHIC,}
\end{equation}
so that at the LHC the average energy of gluons is already as in an 
ideal gas in thermal equilibrium. Only isotropization is needed, and
a rapid thermalization is indeed possible.  An instant thermalization 
would in turn have profound consequences  on {\it e.g.} thermal 
dileptons,  for which a high initial  temperature plays a  
crucial role \cite{RVESA}.\footnote{On the other hand, 
for the thermal dileptons the trouble is the small out-of-equilibrium 
quark-antiquark component of the early parton system.}

Note that our conclusion of the possibility of an almost instant 
thermalization is due to the  small-$x$ enhancement in the HERA 
gluon densities. From the energy/gluon viewpoint it also seems 
that  thermalization for RHIC is going to happen somewhat later. 
Note however, that  above I did not consider isotropization 
of the system at all.  In the simplified picture presented here, 
the transit time of the colliding nuclei, $\tau_{\rm tr}\sim 2R_A/\gamma$,
and the initial parton spread, $\Delta z\sim 1/(xp)\sim 1/p_0$ 
for the partons which will be produced in the mid-rapidity, are neglected. 
Then a Bjorken-like boost-invariant picture is possible, 
and in the central rapidities a proper time $\tau$ is a relevant variable.  
For a more thorough discussion of isotropization, a more detailed 
microscopic space-time picture has to be  specified, 
as done in Refs. \cite{PCM,EW94} 
(see  also the discussion in \cite{EK96,EMW}). 

The fourth observation \cite{EK96} is that initially, at 
$\tau\sim 0.1$ fm/$c$, the net baryon number density in the 
central rapidity unit is very small 
as compared to the gluon  density but {\it larger} than the nuclear matter 
density (0.17 fm$^{-3}$), even though the colliding nuclei are practically 
already far apart,  especially at the LHC where 
$\tau_{\rm tr}\ll \tau_{\rm h}$.
More precisely, we estimate
\begin{equation}
n_{B-\bar B} \equiv \frac{1}{3}(n_q-n_{\bar q})=
\frac{\frac{1}{3}(\bar N_q-\bar N_{\bar q})}{\pi R_A^2\Delta y/p_0} =
\left\{ \begin{array}{ll}
                      0.25\, (0.30)\, {\rm fm}^{-3}, & \mbox{for LHC} \\
                      0.22\, (0.23)\, {\rm fm}^{-3},& \mbox{for RHIC}
                      \end{array}
              \right.
\end{equation}
with the MRSH (MRSD-') parton distributions.
Computing the net baryon-to-entropy ratio by using $s_g = 3.6n_g$ for a 
thermal boson gas gives {\it initially}, at $\tau=0.1$ fm/$c$: 
$(B-\bar B)/S_g \sim 2\cdot10^{-4}$ for LHC, and $\sim 2\cdot10^{-3}$ 
for RHIC.   
We conclude that at the future colliders we are still relatively 
far away from the extreme conditions of the Early Universe, 
where the inverse of the specific entropy is $\sim 10^{-9}$ \cite{UH}.
For the LHC, assuming an instant thermalization of the gluon 
system at $\tau=0.1$ fm$/c$, and an adiabatic evolution thereafter, 
the final entropy can be approximated by the initial entropy of 
gluons \cite{EKR,EK96}. The non-perturbative mechanism for particle 
production will not increase the entropy much but does increase the net 
baryon number. If the non-perturbative contribution to the net baryon number 
production is assumed to be of the same order of magnitude as in the 
current Pb+Pb collisions at SPS, the {\it final} baryon-to-entropy 
ratio for the LHC will be $\sim 10^{-3}$. For the RHIC nuclear collisions, 
thermalization is most likely not as instantaneous, but following 
nevertheless the same line of arguments, and taking into account that 
the non-perturbative component becomes important also for entropy 
production, we estimate $\sim 10^{-2}$ for the final net  
baryon-to-entropy ratio.

\section{Minijets in the BFKL-approach}
Minijet production I have considered above is based on collinear 
factorization, where the perturbative partonic cross sections are 
factorized at a momentum scale $Q\sim p_{\rm T}$
from the parton distributions with nonperturbative input. 
Next, I will discuss an additional 
mechanism for minijet and transverse energy production, 
where factorization is not used.

The small-$x$ rise in the structure function $F_2(x,Q^2)$ observed 
at HERA $Q^2>1.5$ GeV$^2$ \cite{HERA93} 
can be explained by the leading  $\log (Q^2)$ DGLAP-evolution 
\cite{DGLAP} and also by the leading $\log (Q^2)\log(1/x)$ 
evolution \cite{BF}. 
Also a power-like behaviour, $F_2\sim x^{-\delta}$, expected in 
the leading $\log(1/x)$ BFKL-approach \cite{BFKL}, does 
not contradict the data. In the following, let us
assume that the small-$x$ increase is entirely due to the 
BFKL-physics. Then, with this assumption, we will study what 
is the  {\it maximum} transverse energy deposit in the central rapidity 
unit due to the minijets emitted from a BFKL-ladder 
in the LHC nuclear collisions. 
At RHIC the BFKL-minijets are not expected to contribute in any 
significant manner because the BFKL-enhancement takes place 
only at $x\lsim 0.01$. Therefore, this latter part of my talk,
which is based on Ref. \cite{ELR96}, will 
be relevant only for the LHC.

It is instructive to start from a case of fully inclusive minijets
with two tagging jets separated by a wide rapidity gap, as studied by 
Mueller and Navelet \cite{MN87}. The (summed) subprocess is also 
shown  Fig. 1a, where the incoming partons have momentum fractions $x_a$ 
and $x_b$, the tagging jets rapidites $y_a$ and $y_b$ ($y_a\gg y_b$)
and transverse momenta ${\bf k}_{a{\rm T}}$ and ${\bf k}_{b{\rm T}}$,
respectively. Between the tagging jets there are $n$ gluons emitted,
labeled by 1...$n$. 
Thus each final state is described by a Feynman graph with 
2 incoming and $n+2$ outgoing on-shell gluons. The colour singlet hard 
BFKL-pomeron ladder 
arises when these Feynman graphs are squared and summed.
In the kinematic region we will be interested in, 
the rapidities are strongly ordered, $y_a\gg y_1\gg ...\gg y_n\gg y_b$,
but the transverse momenta are not, ${\bf k}_{{\rm T}i}\sim{\bf k}_{{\rm T}j}$.
Then only the transverse degrees of freedom  of the 
momenta of the virtual legs become important. The tagging jets of Fig. 1a 
have transverse momenta at a perturbative scale, so that one may use 
collinear factorization to write the cross section down as:
\begin{equation}
\frac{d\sigma}{d^2{\bf k}_{a{\rm T}}d^2{\bf k}_{b{\rm T}}dy_ady_b} =
\sum^{\infty}_{n=0} x_ag(x_a,\mu^2) x_bg(x_b,\mu^2)
\int \prod_{i=1}^n\frac{dy_i}{4\pi}\frac{d^2{\bf k}_{i{\rm T}}}{(2\pi)^2}
\frac{\overline {|M_{a1...nb}|^2}}{16\pi^2\hat s^2}
\delta^{(2)}(\sum_{j=0}^{n+1}{\bf k}_{j{\rm T}}),
\label{fact}
\end{equation}
where only gluons are considered.
The strong rapidity ordering simplifies the momentum fractions to
$ x_a \approx\frac{k_{a{\rm T}}}{\sqrt s}{\rm e}^{y_a}$ and 
$  x_b  \approx\frac{k_{b{\rm T}}}{\sqrt s}{\rm e}^{-y_b}$,
and the parton densities factor out of the sum.

For the process $gg\rightarrow gg$ the leading contribution in the 
large $\hat s/\hat t$ limit comes from the $t$-channel amplitude. 
In a physical gauge, this amplitude is also gauge invariant up to the 
subleading terms. The matrix element $M_{a1...nb}$ consists then of the 
following building blocks: 
In the leading $\hat s/\hat t$ approximation, in a physical gauge
and with the strong rapidity ordering, each gluon can be 
regarded as emitted from an effective non-local Lipatov-vertex,
where bremsstrahlung from initial and final legs and emission from the 
exchanged gluon are summed. These are described by the black blobs in 
Fig 1. Also the propagators are effective ones since they are 
exponentiated (Reggeized) after computing the leading virtual corrections
to the $t$-channel gluon exchange. The effective propagators are drawn 
by thicker (vertical) lines in Fig. 1. Original references, detailed 
discussion and derivation of these concepts can be found in the useful 
lecture notes by Del Duca \cite{DELDUCA}.

Next, we square each matrix element $M_{a1...nb}$, and due to the strong 
ordering in rapidities, colour singlet ladders with $n+2$ rungs are formed. 
The colour algebra can be performed by summing (averaging) over the final 
(initial) state colours, and the polarization sums can be done. With help of
{\it e.g.} Laplace-transformation (in $y_a-y_b$), the rapidity integrals 
can be disentangled. Finally, by summing over $n$, one obtains an iterative 
integral equation, the inhomogeneous BFKL-equation \cite{BFKL,MN87} 
(see also \cite{DELDUCA}), which describes an addition of one rung into 
the colour-singlet hard pomeron ladder. The BFKL-ladder is denoted by 
$f({\bf q}_{\rm T}, {\bf k}_{\rm T}, y_a-y_b)$
in Fig 1a.  The cross section (\ref{fact}) then becomes:
\begin{equation}
\frac{d\sigma}{d^2{\bf k}_{a{\rm T}} d^2{\bf k}_{b{\rm T}} dy_a dy_b} =
x_ag(x_a,\mu^2)\,x_bg(x_b,\mu^2) \,
\frac{4N_{\rm c}^2\alpha_s^2}{N_{\rm c}^2-1}   \,
\frac{1}{k_{a{\rm T}}^2}     \,
2f({\bf q}_{\rm T}, {\bf k}_{\rm T}, y_a-y_b)  \,
\frac{1}{k_{b{\rm T}}^2},
\label{1LADDER}
\end{equation}
where 
${\bf q}_{\rm T} = {\bf -k}_{a{\rm T}}$ and
${\bf k}_{\rm T} = {\bf k}_{b{\rm T}}$.
If all the virtual corrections and the real emissions are neglected, 
the ladder reduces into $2f({\bf q}_{\rm T},{\bf k}_{\rm T},y)\rightarrow
\delta^{(2)}({\bf q}_{\rm T}-{\bf k}_{\rm T})$, and the Born limit for 
the two jets separated by a large rapidity interval is recovered 
\cite{DELDUCA}.

\begin{figure}[tb]
\vspace{-1cm}
\centerline{\hspace*{-11cm} \epsfxsize=10cm\epsfbox{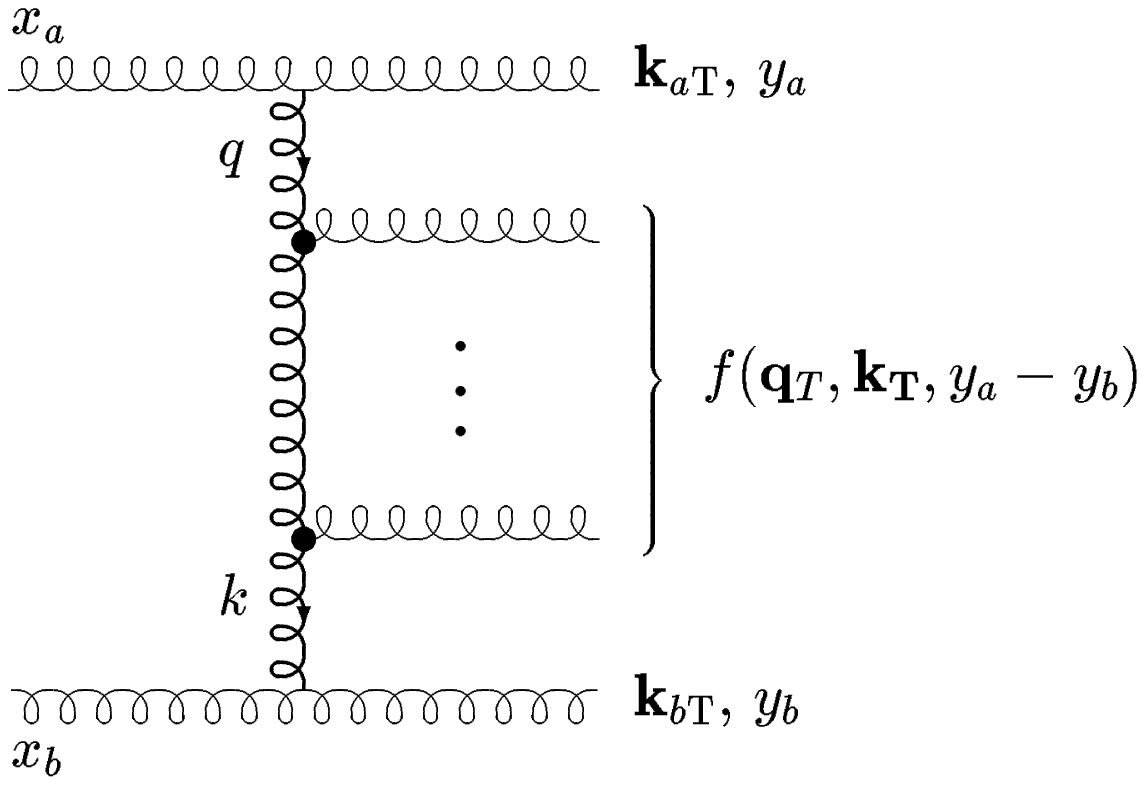}}
\vspace*{-16cm}
\centerline{\hspace*{4cm} \epsfxsize=10cm\epsfbox{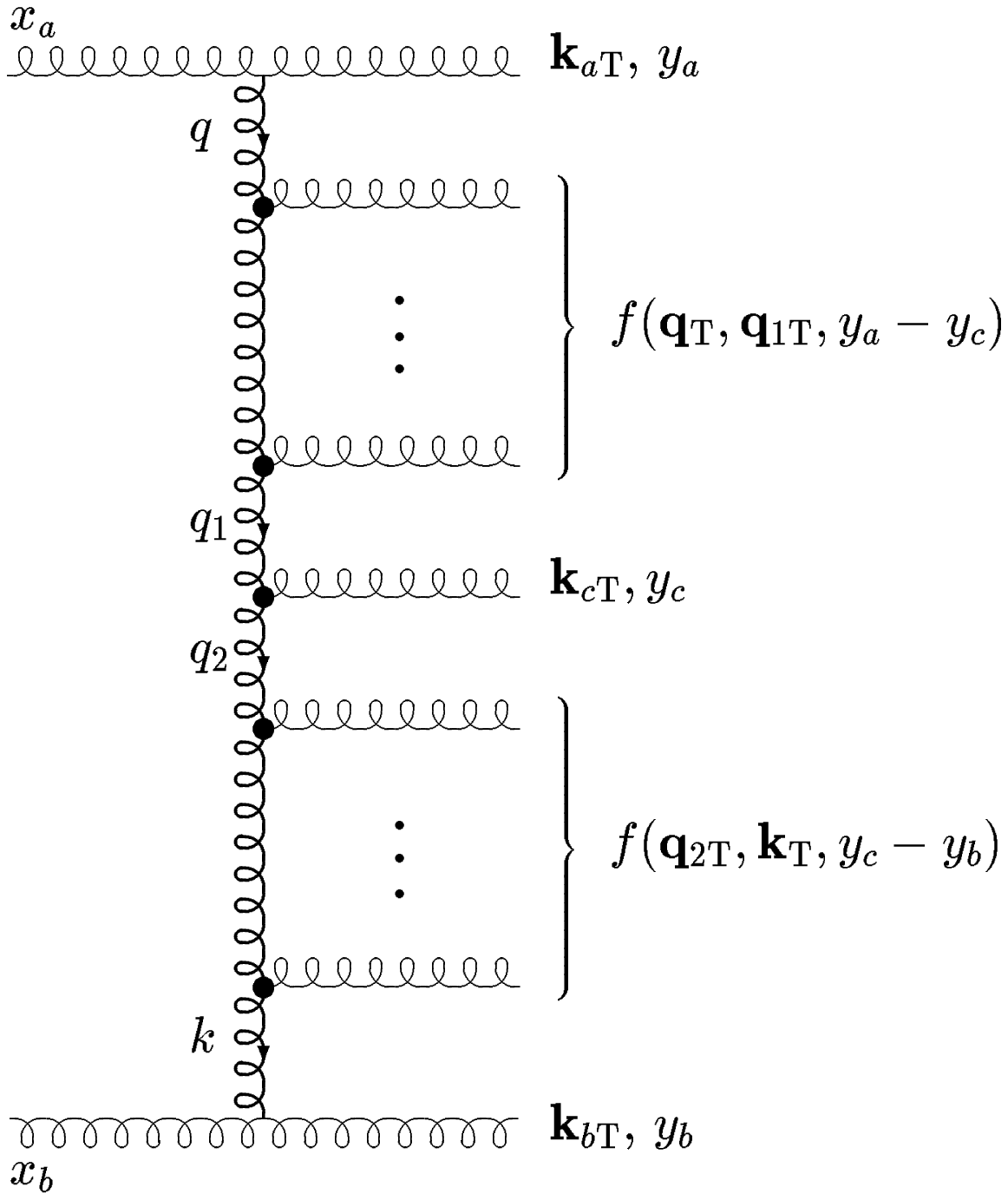}}
\vspace*{-5cm}
\centerline{{\small\bf (a)}\hspace*{6.5cm} {\small\bf (b)}\hspace*{3.5cm}}
\caption[a]{
{\footnotesize {\bf (a)} 
The BFKL ladder in fully inclusive minijet production
between the two tagging jets $a$ and $b$ \cite{MN87,GLRpr,GLRpl}. 
{\bf (b)} 
Fixing one step of the ladder ($c$) creates a ladder 
on each side of the minijet $c$ \cite{DPT}.}
}
\la{graphs1}
\end{figure}

Let us then  study the case with tagging jets further by fixing 
one step of the ladder, as shown in Fig. 1b. It is straightforward
to sum the graphs with gluon emissions between the tagging jet $a$ and the 
fixed minijet $c$, and, between the minijet $c$ and the tagging jet $b$. 
This creates a ladder on each side of the fixed rung. Especially, we 
learn that a generic factor $\alpha_s N_{\rm c}/(\pi^2 k_{c{\rm T}}^2)$, 
which includes the phase-space factor and contraction of the two Lipatov 
vertices associated with the step $c$, arises from fixing the the rung $c$.
The cross section  becomes \cite{DPT}
\begin{eqnarray}
\frac{d\sigma} {d^2{\bf k}_{a{\rm T}} d^2{\bf k}_{b{\rm T}}
d^2{\bf k}_{c{\rm T}} dy_a dy_b dy_c}  =
x_ag(x_a,\mu^2)\,x_bg(x_b,\mu^2) \
\frac{4N_{\rm c}^2\alpha_s^2}{N_{\rm c}^2-1} \,
\frac{\alpha_s N_{\rm c}}{\pi^2}       \,
\frac{1}{k_{c{\rm T}}^2}            \,
\int d^2{\bf q}_{1{\rm T}}d^2{\bf q}_{2{\rm T}}\cdot \nonumber
\\\nonumber
\\
\cdot\,\delta^{(2)}({\bf k}_{c{\rm T}}-
{\bf q}_{1{\rm T}}+{\bf q}_{2{\rm T}})\,
\frac{2f( {\bf k}_{a{\rm T}}, {\bf q}_{1{\rm T}}, y_a-y_c)}{k_{a{\rm T}}^2} \,
\frac{2f( {\bf q}_{2{\rm T}}, {\bf k}_{b{\rm T}}, y_c-y_b)}{k_{b{\rm T}}^2}.
\label{2LADDER}
\end{eqnarray}

\begin{figure}[tb]
\vspace*{-3.5cm}
\centerline{ \epsfxsize=10cm\epsfbox{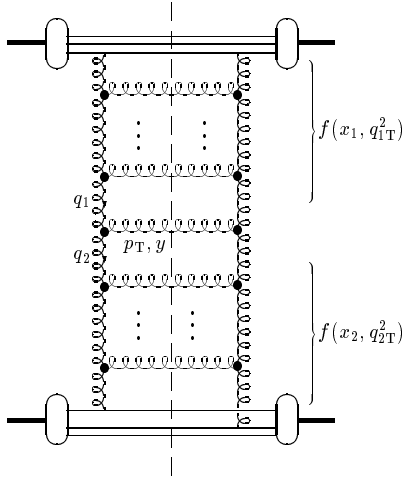}}
\vspace*{-4.5cm}
\caption[a]{{\footnotesize
Minijet production without the tagging jets requires an 
introduction of unintegrated parton distributions $f(x,q_{{\rm T}}^2)$
\cite{GLRpl}.}
}
\la{graph2}
\end{figure}

Our goal is to study the leading BFKL minijet production mechanism 
which is  $\sim\alpha_s$. As illustrated in Fig. 2, we therefore 
relax the requirement of having tagging jets. Then coupling of 
the pomeron ladder to the hadron becomes essentially non-perturbative  
and a form-factor, or, rather, a parton distribution, will be needed. 
Also, now that we do not require any tagging jets, we have to give up 
collinear factorization. We do not have any perturbative Born limit 
to compare with, either. Therefore, the best we can do is to adopt 
the procedure for deep inelastic scattering (DIS) in \cite{AKMS}, 
where an addition of each rung into the pomeron ladder 
between the two hadrons or nuclei is expected to be described by the 
{\it homogeneous} BFKL equation for the unintegrated gluon density
$f(x,q_{\rm T}^2)$,
\begin{equation}
-x\frac{\partial f(x,q_{\rm T}^2)}{\partial x} =
\frac{\alpha_s N_{\rm c}}{\pi}q_{\rm T}^2
\int_{0}^{\infty}
\frac{dq_{1{\rm T}}^2}{q_{1{\rm T}}^2}
\bigg[
\frac{f(x,q_{1{\rm T}}^2)-f(x,q_{\rm T}^2)}{|q_{\rm T}^2-q_{1{\rm T}}^2|} +
\frac{f(x,q_{\rm T}^2)}{\sqrt{q_{\rm T}^4+4q_{1{\rm T}}^4}}
\bigg].
\label{HBFKL}
\end{equation}
Normalization for this scale-invariant equation is given by
the gluon distributions 
\begin{equation}
xg(x,Q^2) = \int^{Q^2}\frac{dq_{\rm T}^2}{q_{\rm T}^2}f(x,q_{\rm T}^2),
\end{equation}
determined from experimental input \cite{AKMS}.

By using the knowledge of the factor arising from fixing one rung of the 
BFKL-pomeron ladder, the inclusive minijet cross section from the 
BFKL-ladder can now be written down as \cite{GLRpl,ELR96}
\begin{equation}
\frac{d\sigma^{\rm jet}}{d^2{\bf p}_{\rm T}dy} =
K_N
\frac{\alpha_sN_{\rm c}}{\pi^2} \,
\frac{1}{p_{\rm T}^2} \,
\int d^2{\bf q}_{1{\rm T}}d^2{\bf q}_{2{\rm T}} \,
\delta^{(2)}({\bf p}_{\rm {\rm T}}-{\bf q}_{1{\rm T}}+{\bf q}_{2{\rm T}}) \,
\frac{f(x_1,q_1^2)}{q_{1{\rm T}}^2} \,
\frac{f(x_2,q_2^2)}{q_{2{\rm T}}^2}
\label{BFKLmini}
\end{equation}
where $p_{\rm T}$ and $y$ are the transverse momentum and the rapidity
(in the hadron CMS) of the minijet. From  momentum
conservation and multi-Regge kinematics the momentum fractions become
$ x_{1(2)} \approx \frac{p_{\,\rm T}}{\sqrt s}{\,\rm e}^{\,\pm y}$.
Due to the fact that in this case we do not have an ``external'' hard
probe like the virtual photon with an associated quark box as
in DIS, nor an on-shell Born cross section to relax into, we cannot
determine the overall dimensionless normalization constant $K_N$ exactly.
However, we are able to fix the {\it slope} of the minijet
$p_{\rm T}$-distribution, which will be sufficient for
estimating the upper limit of transverse energy production from the 
BFKL-ladder.

The minijet cross section of Eq. (\ref{BFKLmini}) is shown in  Fig. 3 
\cite{ELR96}. In the BFKL-computation we have used the unintegrated 
gluon densities compatible with the small-$x$ rise in the set MRS\-D-' 
\cite{MRSD}. For comparison, the more traditional (CFLTLO) minijet 
cross sections, discussed in the first half of the talk, are also 
shown with the MRSD-' parton distributions. The NLO jet analysis 
\cite{EKS,EWatHPC} indicates that LO+NLO calculation with collinear 
factorization reproduces the measured jet cross sections well.
Therefore, at $p_{\rm T}\gsim 5$ GeV, the BFKL-minijet contribution 
should be less than the collinearly factorized. We can thus argue 
that $K_N\lsim 1$. 

\begin{figure}[tb]
\vspace*{4.0cm}
\centerline{ \epsfxsize=8cm\epsfbox{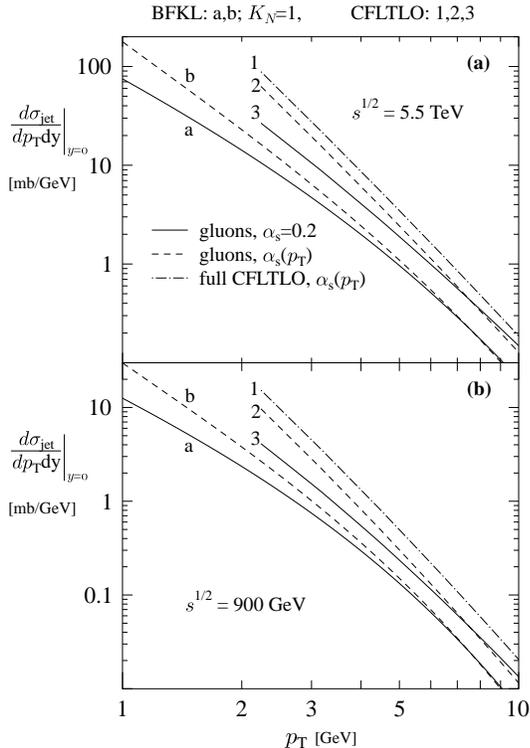}\hspace*{1cm}}
\vspace*{-2.5cm}
\caption[a]{{\footnotesize
The minijet cross sections at $y=0$ as functions of transverse momentum 
$p_{\rm T}$ at $\sqrt s=5.5$ TeV from \cite{ELR96}.
(panel {\rm a}) and $\sqrt s = 900$ GeV (panel \rm {\bf b}) \cite{ELR96}. 
The curves 1, 2 and 3 are predicted in the approach of collinear 
factorization, leading twist and lowest order (CFLTLO) pQCD, and 
MRSD-' parton distributions. The curves a and b are the 
BFKL-minijets from Eq. (\ref{BFKLmini}) with and without running coupling. 
}} 
\la{dsdpT} 

\end{figure}

The transverse energy production due to the minijets from 
the BFKL-ladders at $|y|\le0.5$ in $AA$ collisions can now be 
estimated \cite{ELR96} from
\begin{equation}
\frac{d\bar E_{\rm T}^{\rm BFKL}}{dy}{\bigg|}_{y=0} =
T_{AA}({\bf b})
\int_{p_0^{\rm BFKL}} dp_{\rm T}
p_{\rm T} \frac{d\sigma^{\rm jet}}{dp_{\rm T}dy}{\bigg|}_{y=0}.
\label{BFKLET}
\end{equation}
The coherence of the BFKL ladder is broken when we fix a rung, 
and the cross section diverges at $p_{\rm T}\rightarrow 0$.
A cut-off is, unfortunately, needed also in the BFKL case.
The saturation of the CFLTLO-minijet cross section in the LHC Pb-Pb 
collisions (as considered in the first half of the talk), implies 
that the BFKL-cross section should not grow much larger than the 
curve 2 in Fig. \ref{dsdpT}. Therefore, we do not trust
the BFKL-computation with $K_N\sim 1$ below 
$p_{\rm T}\le p_0^{\rm BFKL}\sim 1$ GeV.
With these values, we find for central Pb-Pb collisions at the LHC,  
$\bar E_{\rm T}^{\rm BFKL} = 3060$ GeV with fixed $\alpha_{\rm s}=0.2$, and,
4940 GeV with ({\it ad hoc}) running $\alpha_{\rm s}(p_{\rm T})$.
Comparing these numbers to the value 15330 GeV in Table 1b for gluons,
we see that the BFKL-contribution is at most a few 10\% correction 
to the results in \cite{EKR,EK96}. On the other hand, one should 
perhaps compare the results at the same level of approximations,
(only LO gluons, $\alpha_{\rm s}$ fixed) {\it i.e.} curves 3 and 
a in Fig. \ref{dsdpT}a. Then the two contributions become of similar 
magnitude. In this case, however, the  $p_0$ in the CFLTLO-computation 
should be lower than 2 GeV, and the BFKL contribution would again be 
the smaller one. 

We worked under the assumption that the BFKL-evolution is responsible 
for {\it all} the small-$x$ rise at HERA, {\it i.e.} we studied the 
{\it maximum}  contribution from the kinematical region relevant 
for the hard BFKL-pomeron.  Since the HERA results can be explained 
by the leading  $\log(Q^2)$ and/or the leading $\log(Q^2)\log(1/x)$ 
approximations, the leading $\log(1/x)$-contribution is obviously 
not the dominant mechanism at the present values of $x$. Thus, my 
conclusion is that the BFKL-minijets certainly bridge the way towards 
softer physics at $p_{\rm T}<p_0\sim 2$ GeV, but the initial conditions 
relevant for the early QGP-formation in the LHC nuclear collisions 
are dominantly given by the minijets computed in collinear factorization. 

\bigskip
\noindent{\bf Acknowledgements.} 
The results discussed in this talk are based on Refs.
\cite{EKR,EMW,EK96,ELR96}.
I would like to thank K. Kajantie, A. Leonidov, B. M{\"u}ller, 
V. Ruuskanen and X.-N. Wang for fruitful collaboration.
I also owe special thanks to A. Leonidov and V. Ruuskanen for 
getting our BFKL-study started and finally finished.

\end{document}